\documentclass[runningheads]{llncs}
\usepackage{graphicx}
\usepackage{caption} 
\begin{document}
\title{Extracting gamma-ray information from images with convolutional neural network methods on simulated \mbox{Cherenkov Telescope Array} data}
\titlerunning{Convolutional Neural Networks for CTA Data Reconstruction}
%
\author{\mbox{S. Mangano, C. Delgado, M. Bernardos, M. Lallena, J. J. Rodr\'iguez V\'azquez,} \mbox{for the CTA Consortium\thanks{CTA website: \url{https://www.cta-observatory.org/.}}}}

\institute{CIEMAT - Centro de Investigaciones Energ\'eticas, Medioambientales y Tecnol\'ogicas\\ Av. Complutense, 40, 28040 Madrid, Spain\\ 
\url{http://cta.ciemat.es/}
}

\authorrunning{S. Mangano et al.}
%
%
\maketitle              
\begin{abstract}
The Cherenkov Telescope Array (CTA) will be the world's leading ground-based gamma-ray observatory allowing us to study very high energy phenomena in the Universe. CTA will produce huge data sets, of the order of petabytes, and the challenge is to find better alternative data analysis methods to the already existing ones. Machine learning algorithms, like deep learning techniques, give encouraging results in this direction. In particular, convolutional neural network methods on images have proven to be effective in pattern recognition and produce data representations which can achieve satisfactory predictions. We test the use of convolutional neural networks to discriminate signal from background images with high rejections factors and to provide reconstruction parameters from gamma-ray events. The networks are trained and evaluated on artificial data sets of images. The results show that neural networks trained with simulated data can be useful to extract gamma-ray information. Such networks would help us to make the best use of large quantities of real data coming in the next decades.

\end{abstract}
\section{Introduction}
The ground-based observation of the
very high energy gamma-ray sky (E$_{gamma}>100~GeV$)
has greatly progressed during the last 40 years 
through the use of imaging atmospheric Cherenkov telescopes (IACTs).   
These telescopes aim to detect the
air shower produced by the interaction of a primary cosmic
gamma ray in the Earth's atmosphere. 
Charged air shower particles that travel
at ultra-relativistic speed emit Cherenkov light. 
This Cherenkov light propagates 
to the ground producing a faint pool of Cherenkov light of 
about 120 m in radius. 
The optical mirrors of the telescopes reflect 
the collected Cherenkov light into the focal plane where 
photomultipliers convert light into an electrical signal 
that is digitized and transmitted to record the image.

The image in the camera represents the electromagnetic air shower 
and is used to identify the primary 
cosmic gamma-ray. However, Cherenkov light is not 
only produced by cosmic gamma-rays but also by the more abundant hadronic cosmic rays. These 
massive charged particles arriving from outer space are mostly protons, but they also include heavier nuclei, which are known atoms without their electron shells.
The shape, intensity and orientation of the image 
provides information about the primary cosmic particle type, energy, direction of propagation and depth of first interaction. 

Several different classification and reconstruction techniques  
exist which on one hand discriminate gamma-ray events 
from the more numerous hadron events and on the other hand infer the primary gamma-ray energy and direction.
One of the first developed reconstruction methods~\cite{Hillas}, the so called Hillas parametrization, used 
direction and elliptical shape of the gamma-ray images as the main features 
to discriminate them 
against the hadronic cosmic ray background 
which produces wider and more irregular images.
Later more advanced reconstruction methods with superior 
performance have been developed using machine learning algorithms 
as in the case of random forest~\cite{Randomforest} for the MAGIC~\cite{MAGICpaper} telescope and 
boosted decision trees~\cite{Ohm} for the H.E.S.S.~\cite{HESSpaper} and VERITAS~\cite{VERITASpaper} telescopes. 
A further reconstruction method is to fit the image to results of a fast simulation under the hypotheses 
that the image is an electromagnetic shower~\cite{Mathieu,Holler,impact}.

Recently, several gamma-ray observatories with Cherenkov telescopes started
using  convolutional neural networks (CNNs) for classification and regression problems~\cite{FengLin,Tim,Daniel,Shilon}.
CNNs belong to a class of supervised machine learning techniques
that have achieved impressive results in image processing~\cite{Lecun,Krizhevsky} with little need of 
human intervention in finding significant image features. 
With enough training data CNNs can find patterns in the data that when applied to images 
maximize the gamma-ray reconstruction performance or background rejection. 
In general the study of such machine learning follows always the 
same procedure: 
to start define a data set, then determine a cost function that has to be minimized, next design a neural network 
architecture where computationally efficient changes on adjustable parameters works, and in the end apply 
some sort of stochastic gradient descent to minimize the cost function. 
For an in depth treatment of the literature, see the following references~\cite{book,Lecun2,deeplearningbook}.

The Cherenkov Telescope Array~\cite{CTApaper,CTAsalvi} (CTA) will be the next generation ground-based gamma-ray observatory 
to study very high energy processes in the Universe.
The main goal of CTA is to identify and study high energy gamma-ray sources, including objects such as supernova remnants, pulsars, binary stars and 
active galaxies. The measured fluxes, energy spectra and arrival directions 
of gamma rays will help to find answers to the origin of these high energy particles and provide information on the morphology of the sources. 
Also some more speculative models are investigated, like theories which incorporate the violation of Lorentz invariance
and predict unexpected cosmological effects on gamma-ray propagation, or 
the search of possible signals from annihilating dark matter particles.
CTA is expected to have around one order of magnitude 
improvement in sensitivity in the energy range from $\sim 50$ GeV to $\sim 50$ TeV compared 
to currently operating IACTs. This is due to the fact that CTA has the capability to detect gamma-rays over larger areas than existing observatories. 
CTA will provide whole-sky coverage with an observatory in the Southern Hemisphere (Cerro Paranal, Chile) and an observatory in the Northern Hemisphere (La Palma, Spain). The Southern Hemisphere observatory has a total of 99 telescopes of three different sizes with an area of 4.5 km$^2$ and the Northern Hemisphere observatory has a total of 19 telescopes with an area of 0.6 km$^2$. 
These telescopes will provide a large amount of images
that encode primary particle information and it is essential to 
develop efficient statistical tools to best
extract such information. Moreover both observatories will be equipped with four large size telescopes~\cite{LST}, each with a mirror diameter of about 23 m and a focal length of 28 m. The large size telescope will dominate the performance of the observatory between 20 GeV and 200 GeV and
will be equipped with a 1855 pixels camera with 4.6 degree full field of view.
First real data from such a telescope should be available already in the end of 2018.

In this note, we aim to asses the use of CNNs to discriminate signal from background images and to provide reconstruction parameters from gamma-ray events for the CTA observatory.
To evaluate the performance of the CNNs 
we use official simulated CTA data exploiting the pixel wise information 
of minimally treated images. In contrast to previous mentioned 
existing works, 
we apply for the first time CNNs to simulated CTA data to
reconstruct the gamma-ray parameters.
We focus only on large size telescopes with showers triggered in all four telescopes.
The remainder of this note is organized as follows.
Section~\ref{sec:simulation} gives a short description of data simulation and data selection. In Section~\ref{sec:tensorflow} we present details about specific networks, explain analysis strategy and discuss results, followed in Section~\ref{sec:conclusion} by concluding remarks.

\section{Monte Carlo Simulation and Preselection}\label{sec:simulation}
A Monte Carlo simulation has been used~\cite{CTABernloer,Maier} to produce a large artificial data set\footnote{The simulation data used for this study were extracted from the so called CTA prod-3 data set.} and to examine the performance of different CNN architectures. As presented in~\cite{CTABernloer} the Monte Carlo generated gamma-ray data has been verified against real gamma-ray data from the existing Cherenkov telescopes. For this study the directions of the primary gamma rays and protons are distributed isotropically and extend well beyond the CTA field of view. In particular, for this diffuse emission, no previous knowledge about the true direction of the primary gamma-ray source position is assumed. 
The development of extensive air showers caused by primary gamma-rays and protons including emission of Cherenkov light
is simulated with CORSIKA~\cite{Corsika}. 
The primary particles enter the atmosphere as 
diffuse emission within 10 degrees of the field of view center with an average 
zenith angle of 20 degrees and an average azimuth angle 
of 0 degrees. These events have been produced in the energy range from 3 GeV to 330 TeV for gamma rays and from 4 GeV to 660 TeV for protons.
The distinct energy range values is due to the fact that at the same energy, the Cherenkov photon intensity in a proton shower is smaller than the one produced in a gamma-ray shower. The ratio of Cherenkov photon yield
between gamma-ray shower and proton shower is around two for the selected energy range. In proton showers 
a large fraction of the total energy is carried by hadrons and neutrinos which produce little or no amounts of Cherenkov photons~\cite{Ong}. 
The atmospheric conditions of the La Palma site have been reproduced and 
the response of the telescope is simulated by the sim\_telarray~\cite{simtel} package. The generated camera images of telescopes consisting of calibrated integrated charge and pulse arrival time per pixel is extracted from the simulation using the MARS~\cite{MARS} package.

\begin{figure}[t!]
  \centerline{\includegraphics[width=1.0\linewidth]{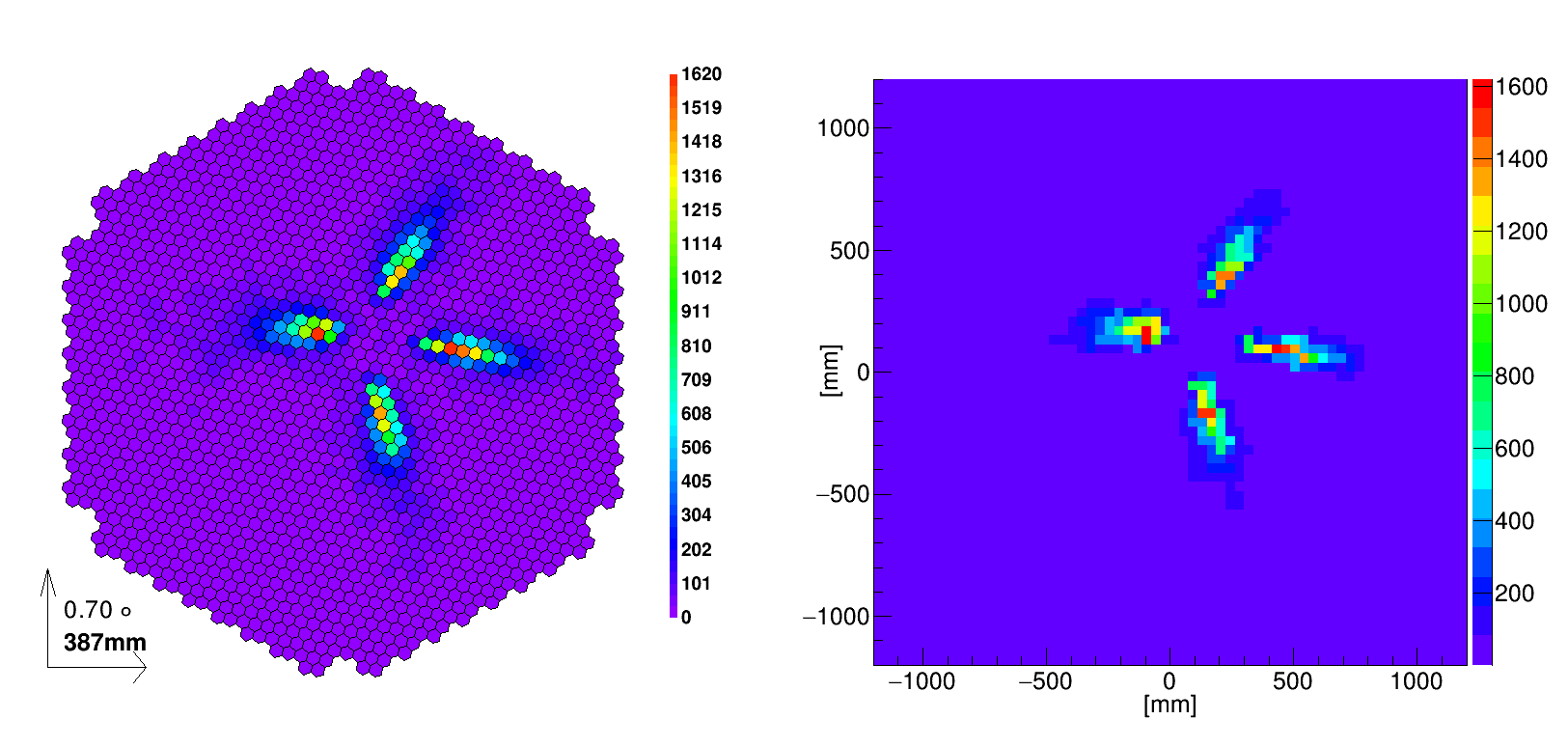}}
\caption[]{Left: Camera pixel intensity of a four combined telescope image of a gamma-ray event in a hexagonal grid with hexagonal pixels. Right: Same event as in the left Figure but as a squared image with squared pixels produced by oversampling technique.}
\label{fig:fourshowers}
\end{figure}

The main aim for IACTs is
to fully reconstruct properties like type, energy, direction and depth of first interaction of the primary particle from the Cherenkov light produced by atmospheric shower. 
The use of more than one telescope significantly improves
the ability to reconstruct these particle properties as the air 
shower can be recorded under different
viewing angles, usually referred to as stereoscopic imaging.
To incorporate this stereoscopic information and reduce the complexity of different numbers of telescopes we select only events that trigger
four large size telescopes. 
To simplify the further analysis we 
combine four images 
into a single image by summing pixel values. 
As CTA images are arranged in a hexagonal grid like the one presented in Figure~\ref{fig:fourshowers} left, whereas the 
CNN framework is 
designed to process only rectangular pixels, some image processing is needed.
A straightforward conversion from the hexagonal (1855 pixels) to squared image (64 $\times$ 64 pixels) is to use an oversampling technique. One such realization is presented in the Figure~\ref{fig:fourshowers} right. 
The CNN has been supplied with such preprocessed integrated charge per pixel images and with labels like primary particle type, energy, direction and depth of first interaction.

The following selection criteria have been used to simplify the reconstruction task.
The incoming directions of the randomized primary particles were selected within a cone with four degrees radius centered on the pointing direction. The impact points of the uniformly distributed primary particles on the ground have been selected within a circle with a radius of \mbox{200 m} around the coordinates of each single telescope. This ensures that the superimposed 
elliptical images do not overlap much. In principle such a selection can be done as a two class classification problem, distinguishing images with small overlap versus large overlap.
However for this study we did not include such a classification selection and we leave this as a future development. 

\section{Convolutional Neural Networks for Simulated Cherenkov Telescope Array Data}\label{sec:tensorflow}
We present results of one CNN that separately classifies signal and background events and a second one that reconstructs parameters of the primary gamma-ray particles.
We use \mbox{TensorFlow~\cite{Tensorflow}} to implement a network architecture handling as input the preprocessed images mentioned in the previous section.
In the following, we give details
about architecture and training of the CNNs and
provide examples of applications to official simulated CTA data.

A typical CNN architecture consist of several successive convolutional layers followed by one or more fully connected layers.
In the first convolutional layer the input image is 
convolved by a filter (also referred as kernel) over a restricted region (also referred as receptive field) producing activation maps. 
The restricted region is in general much smaller than the input images and allows to identify in the first layer simple features, like edges or curves. 
Applying filters on following layers obtain activation maps that represent more and more complex features producing an automated feature extractor.
Such a feature extractor can possibly identify discriminative information in the images that is not fully exploited 
by existing reconstruction algorithms. 

The goal of the CNN is not to achieve good predictions 
on training data examples, but to make good predictions for new examples that are not contained in the training set. 
This requires that the neural network finds the underlying main information in
data and generalizes in a meaningful way.
Various neural network architectures were trained tuning hyperparameters in order to optimize performance on the test set. The performance is given by energy and angular resolution. Once the architecture and hyperparameters are decided, the algorithm is fully automatic.
Due to the large amount of possible parameter combinations, the currently used solution was obtained by random search. Several different sequential architectures with two, four and eight convolution layers combined with one or two fully connected layers with different activation functions and kernel sizes have been tested. Usually neural networks which have larger numbers of parameters may generalize better than neural networks with fewer parameters, 
but larger networks may have an increased overfitting problem and may require longer periods of time in order to complete a calculation than smaller networks.
Even if training of CNNs can take huge computer resources, the finally trained network reconstructs a new event in a short time compared to the training time.
Using simpler architectures over more complex architectures with similar performance reduce reconstruction times and so reduce computing costs. Quicker reconstruction means quicker scientific results, which is better for many scientific objectives such as for example transient phenomena and short-timescale variability searches. 

The selected architecture, which gives reasonable performance in terms of function loss during testing, consists of four convolutional layers with a kernel of 5 $\times$ 5 pixels and with a feature sizes of 32 in each layer.
The convolutional layers and fully connected layers both had exponential linear unit activation for the regression and classification problem. 
In order to account for rotational invariance, the data is augmented artificially with rotated examples (e.g. 0, 120 and 240 degrees)\footnote{One approach producing similar results as the one explained in the text was to use
harmonic networks~\cite{harmonic} to grasp the rotational invariance of the problem.}. However the rotational invariance is only an approximation, since the geomagnetic field actually breaks such a symmetry.
Each convolution layer is followed by a batch normalization layer~\cite{batchnorm} and an average pooling layer~\cite{pooling}, with pool size of two and stride length of two, which reduce the size of images to half in pixels. 
A dropout layer~\cite{dropout} with 80\% to keep the neurons is used during training, whereas  
at the final test time dropout uses all neurons.

The flattened representations from the fourth convolution layer is then followed by a fully connected layer 
of 256 parameters with the same activation functions used in the previous layers. Finally we apply a sigmoid activation function for 
probabilistic predictions in the classification problem and no activation function in the regression problem.

\begin{figure}[t!]
    \centering
    \begin{minipage}[t]{0.45\textwidth}
        \centering
        \includegraphics[width=1.\textwidth]{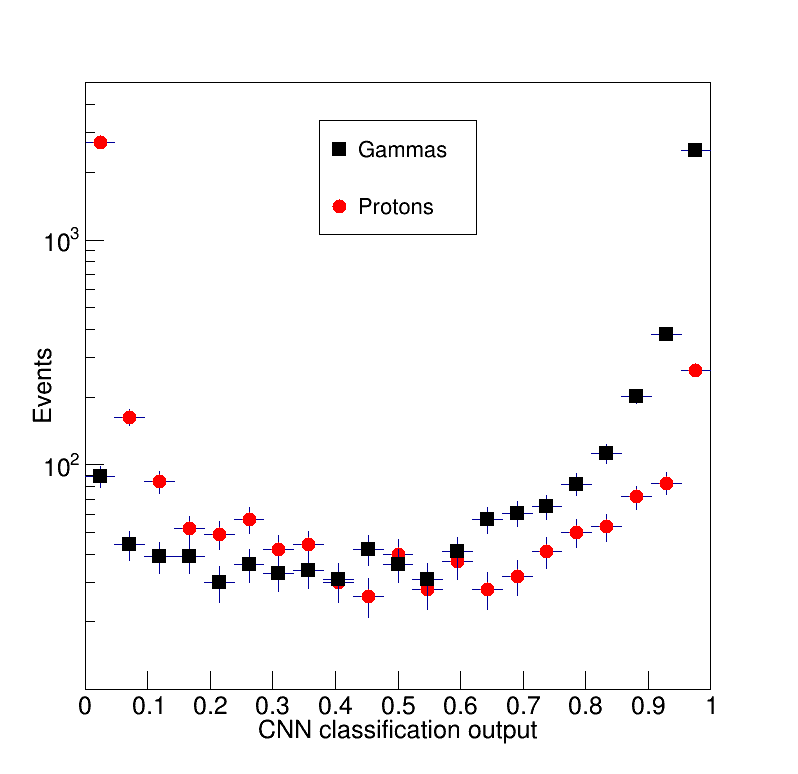} 
        \caption{The CNN gamma-ray and proton classification output for events using an independent test set.}
        \label{fig:detectinggammavsproton}
    \end{minipage}\hfill
    \begin{minipage}[t]{0.45\textwidth}
        \centering
        \includegraphics[width=1.\textwidth]{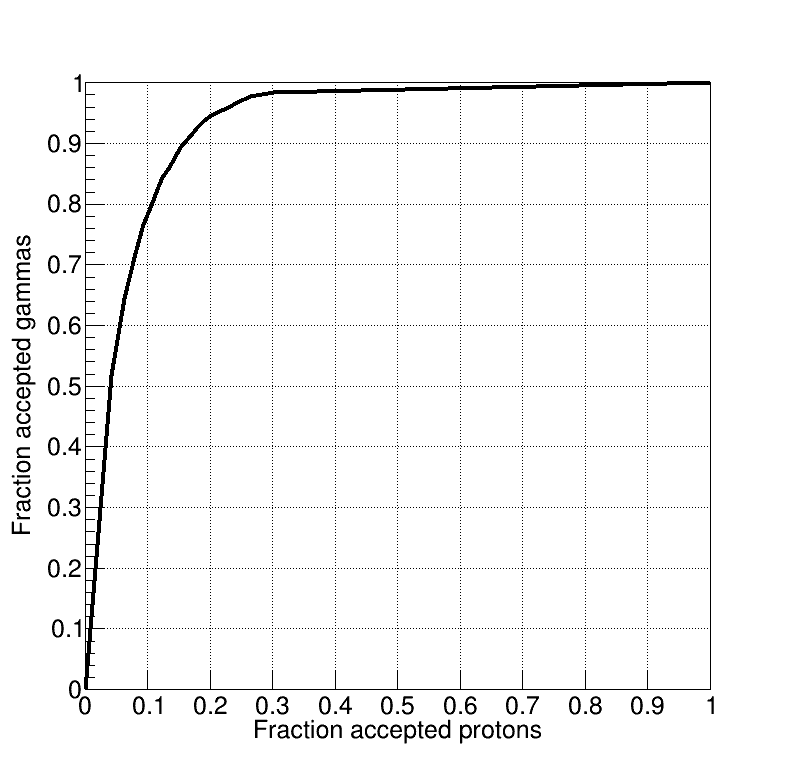} 
        \caption{ROC curve for simulated true energy for gamma-ray and proton events above \mbox{10 GeV}.}
        \label{fig:roccurve}
    \end{minipage}
\end{figure}

The initialization scheme used for the parameters is commonly referred to as the
Xavier initialization~\cite{Xavier}. The cost function for the classification problem is cross entropy and for the regression problem is mean squared error. 
Backpropagation~\cite{rumelhart} is explicitly used to minimize the cost function by adapting parameters using a gradient descent optimization algorithm~\cite{optalgo}.
Training proceeds by optimizing the cost function with L2 regularization and learning rate decay using the Adam algorithm~\cite{adam}.
At each training step, we select a random sample of simulated data with batch size of 256 and use them to optimize the network parameters.
The models were trained on a cluster with Tesla K80 GPUs.
The data set for the classification problem consists of the same number of gamma-ray and proton events 
with about 24000 simulated events and for the regression problem consists of about 40000 gamma-ray events.  
The data was randomly divided into two sets: a training set (80\%) and a test set (20\%).

After having trained the CNN for the classification problem, the classifier is tested with an independent test set of gamma-ray and proton events. 
As an example, Figure~\ref{fig:detectinggammavsproton} shows the result of the 
classification of this test set with the trained CNN, representing the classification power of the CNN approach in terms of gamma-ray and proton separation for events above 10 GeV. 
To illustrate the general performance of our binary classification problem, we use the receiver operator characteristic (ROC) curve shown in Figure~\ref{fig:roccurve}. The ROC curve is a graphical plot that illustrates the true positive rate versus the false positive rate for each possible discrimination value.

\begin{figure}[t!]
    \centering
    \begin{minipage}[t]{0.45\textwidth}
        \centering
        \includegraphics[width=1.\textwidth]{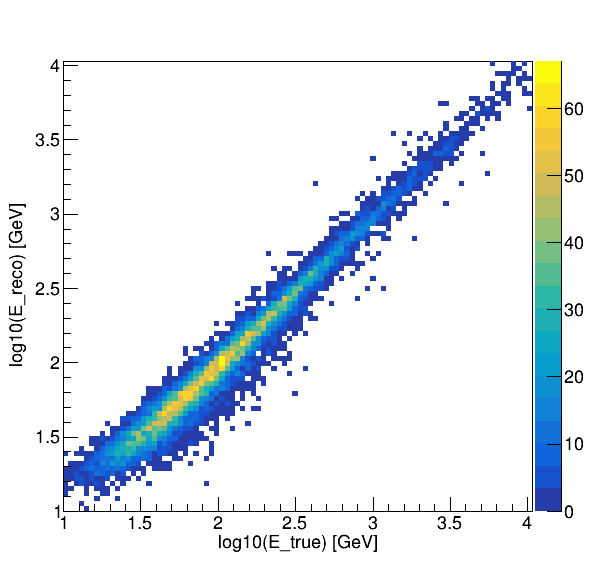} 
        \caption{Reconstructed energy as function of simulated true energy for only diffuse gamma-ray events using a separate CNN than the one used for classification.}
        \label{fig:detectingtruevsrec}        
    \end{minipage}\hfill
    \begin{minipage}[t]{0.45\textwidth}
        \centering
        \includegraphics[width=1.\textwidth]{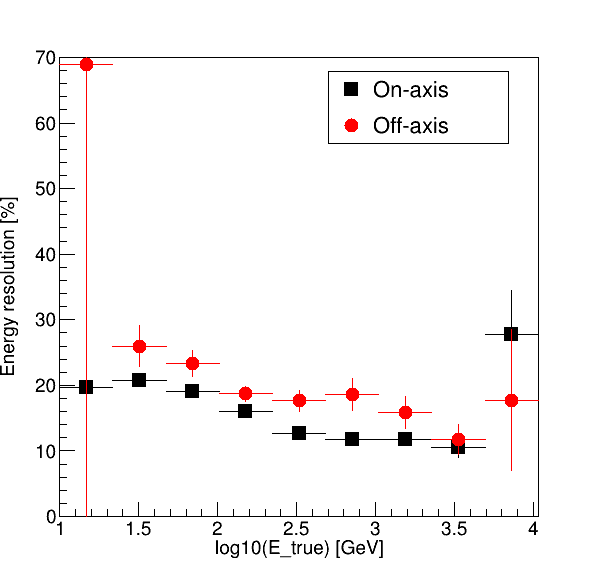}
        \caption{Energy resolution as a function of simulated true energy for two different data sets of diffuse gamma-ray events. The on-axis (off-axis) points represent the resolution of diffuse gamma-ray events with angles with respect to the field of view center of less (more) than two degrees.}
        \label{fig:energyres}
    \end{minipage}
\end{figure}

We trained separate dedicated CNNs to estimate gamma-ray energies, direction and depth of first interaction. 
The trained networks for the regression problem are able to 
reproduce the simulated energy of the events as seen in the Figure~\ref{fig:detectingtruevsrec}, where reconstructed energy as a function of true energy is presented. Figure~\ref{fig:energyres} shows the energy resolution as a function of the true energy of our CNN for two different data sets. The on-axis and off-axis data set represent the energy resolution of diffuse gamma-ray events with angles with respect to the field of view center of less and more than two degrees, respectively. The energy resolution is defined as the one standard deviation of a Gaussian function fit of the distribution of the difference between true and reconstructed energy divided by true energy for a given energy range. The expected energy resolution performance of CTA~\cite{Maier} based on combination of Hillas parametrization and multivariate classification methods is slightly better with about an energy resolution of 9\% at 300 GeV. Table~\ref{tab:table1} compares the energy resolution of the baseline algorithm with the results of this work for three distinct energy bins. However these numbers hava to be taken with care as such comparison are dependent on the differences in data sample like diffuse and point like emission, data selection, number of telescopes and selected strategy.

\begin{table}[b!]
\begin{center}
\begin{tabular}{|c|c|c|c|c|}
\hline
\multicolumn{1}{|c}{} &
      \multicolumn{2}{|c|}{Energy resolution [\%]} &
      \multicolumn{2}{c|}{Angular resolution [deg]} \\
\hline
Simulated true energy  & Baseline algorithm & This work & Baseline algorithm & This work  \\
\hline
30 GeV   & 25$\pm$0.5 & 21$\pm$0.4 & 0.26$\pm$0.005 & 0.26$\pm$0.01 \\
300 GeV  & 9$\pm$0.5 & 13$\pm$0.4 & 0.09$\pm$0.005 & 0.10$\pm$0.005 \\
3000 GeV & 7$\pm$0.5 & 11$\pm$1.6 &  0.05$\pm$0.005 & 0.08$\pm$0.01 \\
\hline
   \end{tabular}
\caption{Comparison of energy resolution and angular resolution for three simulated true energy bins for baseline CTA reconstruction algorithm and the CNN reconstruction presented in this work. Lower values implies better resolution. For the energy and angular resolution only statistical uncertainties are shown.}
\label{tab:table1}
  \end{center}
\end{table}

\begin{figure}[t!]
    \centering
    \begin{minipage}[t]{0.45\textwidth}
        \centering
        \includegraphics[width=1.\textwidth]{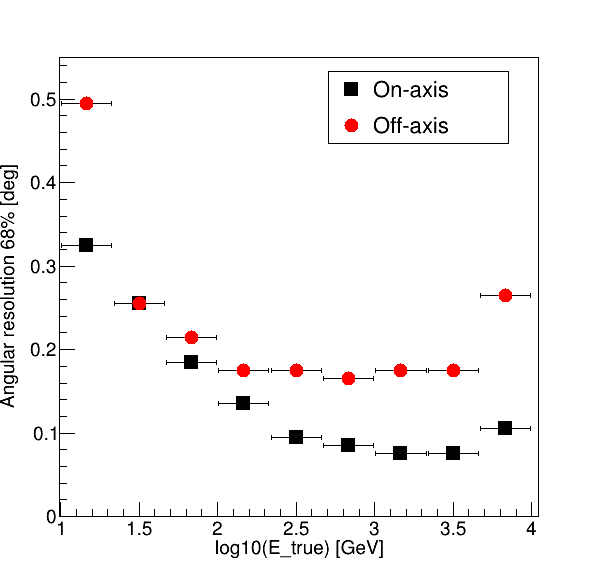}
        \caption{Angular resolution as a function of simulated true energy for two different data sets of diffuse gamma-ray events. The on-axis (off-axis) points represent the resolution of diffuse gamma-ray events with angles with respect to the field of view center of less (more) than two degrees.}
        \label{fig:angularres}       
    \end{minipage}\hfill
    \begin{minipage}[t]{0.45\textwidth}
      \centering
      \includegraphics[width=1.\textwidth]{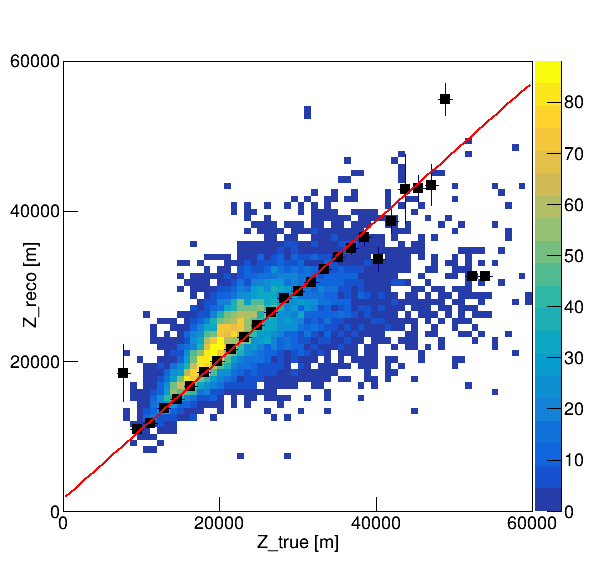}
      \caption{Reconstructed depth of first interaction as a function of simulated true depth of first interaction using the same CNN architecture as for reconstruction of primary particle direction.}
        \label{fig:ztruevsrec}    
    \end{minipage}
\end{figure}

The directional reconstruction performance as a function of true energy of the CNN is given in the Figure~\ref{fig:angularres} for the two different on-axis and off-axis data sets. As can be seen from the Figure~\ref{fig:angularres} the on-axis angular resolution is better than the off-axis one.  
The angular resolution is defined as the angular offset, relative to the true
gamma-ray direction, within which 68\% of the 
gamma-ray events are reconstructed. 
The angular resolution
for a point like emission for the CTA baseline algorithm is about 0.09 degrees at 300 GeV. Table~\ref{tab:table1} shows the angular resolution of the baseline algorithm and the results of this work for three distinct energy bins.

Finally, in contrast to energy and directional primary particle reconstruction, the reconstruction of depth of first interaction of the primary particle is not used in many analyses, although depth of first interaction is useful to separate lepton from gamma-ray initiated showers. This quantity can be difficult to reconstruct if the number of triggered telescopes is small. Moreover, the algorithm~\cite{JulianSitarek} used to reconstruct this variable 
needs knowledge about the physics interaction and detector response. In contrast 
CNN algorithm needs no additional physical knowledge except what is in the simulation and we use the same CNN architecture as for directional reconstruction.
Figure~\ref{fig:ztruevsrec} shows the reconstructed depth of first interaction as a function of true depth of that interaction. A clear correlation is seen suggesting that the height of first interaction can be estimated automatically without any further changes on the CNN architectures. 

In this study we did not exploit all the information as images should be separated according to individual telescopes. We should also take advantage of including all relevant telescope types and the timing information. 
It has been shown~\cite{Stamatescu} that the primary particle information as well as the background rejection can be significantly improved by using timing knowledge.

\section{Summary and Conclusion}\label{sec:conclusion}
The aim of this work is to investigate a deep learning technique for atmospheric Cherenkov telescopes classification and primary particle 
parameter estimation. 
The approach of the work is to treat gamma-ray detection 
as a two class classification problem (gamma-ray versus proton events) as well as to reconstruct gamma-ray shower parameters
and solve it with supervised learning methods.

Promising CNN results have been found and a first comparison 
to previously published baseline algorithm can be made. 
The main advantages of CNN over existing algorithm is that there is 
little need of specialized physics knowledge 
with minimal preprocessing of data. 
Although the results are still not as good as a existing model based algorithms, CNN have simpler implementation requiring no detailed physics assumptions. 

Further analysis on network architecture and image preprocessing 
is needed to improve reconstruction results. 
Specifically our method does not exploit the full information as images should be separated according to individual telescopes. 
We leave the study for a more general CNN taking into account of more sophisticated approaches, 
like use hexagonal symmetric features, include timing information and use all telescope types 
for the future work.  All these steps are required
to add more complexity and generalize our analysis in order 
to provide a more performing CNN for upcoming CTA data.

\section*{Acknowledgments}
We gratefully acknowledge the support of the project (\mbox{reference} AYA2014-58350-JIN) funded by MINECO through the young scientist call (year 2014).
This work is partially supported by the Maria de Maeztu Units of Excellence Program (\mbox{reference} MDM-2015-0509).
We also gratefully acknowledge financial support from the agencies and organizations listed here: http://www.cta-observatory.org/ consortium\_acknowledgments. 
This research shows in Table~\ref{tab:table1} the CTA energy and angular resolution values provided by the CTA Consortium and Observatory, see http://www.cta-observatory.org/science/cta-performance/ (version prod3b-v1) for more details. 
We want to thank K. Bernl\"ohr for the simulation CTA prod-3 data set, which was carried out at the Max Planck Institute for Nuclear Physics in Heidelberg.
This work was conducted in the context of the CTA \mbox{Analysis} and Simulation Working Group and this paper has gone through internal review by the CTA Consortium.

%
%
%
%

\end{document}